\documentclass[12pt,epsf]{article}
\usepackage{graphicx,amsmath,amssymb,booktabs,enumerate}
\usepackage{subfigure}
\begin{document}

\def\a{\alpha}
\def\b{\beta}
\def\c{\varepsilon}
\def\d{\delta}
\def\e{\epsilon}
\def\f{\phi}
\def\g{\gamma}
\def\h{\theta}
\def\k{\kappa}
\def\l{\lambda}
\def\m{\mu}
\def\n{\nu}
\def\p{\psi}
\def\q{\partial}
\def\r{\rho}
\def\s{\sigma}
\def\t{\tau}
\def\u{\upsilon}
\def\v{\varphi}
\def\w{\omega}
\def\x{\xi}
\def\y{\eta}
\def\z{\zeta}
\def\D{\Delta}
\def\G{\Gamma}
\def\H{\Theta}
\def\L{\Lambda}
\def\F{\Phi}
\def\P{\Psi}
\def\S{\Sigma}

\def\o{\over}
\def\beq{\begin{eqnarray}}
\def\eeq{\end{eqnarray}}
\newcommand{\gsim}{ \mathop{}_{\textstyle \sim}^{\textstyle >} }
\newcommand{\lsim}{ \mathop{}_{\textstyle \sim}^{\textstyle <} }
\newcommand{\vev}[1]{ \left\langle {#1} \right\rangle }
\newcommand{\bra}[1]{ \langle {#1} | }
\newcommand{\ket}[1]{ | {#1} \rangle }
\newcommand{\EV}{ {\rm eV} }
\newcommand{\KEV}{ {\rm keV} }
\newcommand{\MEV}{ {\rm MeV} }
\newcommand{\GEV}{ {\rm GeV} }
\newcommand{\TEV}{ {\rm TeV} }
\def\diag{\mathop{\rm diag}\nolimits}
\def\Spin{\mathop{\rm Spin}}
\def\SO{\mathop{\rm SO}}
\def\O{\mathop{\rm O}}
\def\SU{\mathop{\rm SU}}
\def\U{\mathop{\rm U}}
\def\Sp{\mathop{\rm Sp}}
\def\SL{\mathop{\rm SL}}
\def\tr{\mathop{\rm tr}}

\def\IJMP{Int.~J.~Mod.~Phys. }
\def\MPL{Mod.~Phys.~Lett. }
\def\NP{Nucl.~Phys. }
\def\PL{Phys.~Lett. }
\def\PR{Phys.~Rev. }
\def\PRL{Phys.~Rev.~Lett. }
\def\PTP{Prog.~Theor.~Phys. }
\def\ZP{Z.~Phys. }


\baselineskip 0.7cm

\begin{titlepage}

\begin{flushright}
UT-12-05
\end{flushright}

\vskip 1.35cm
\begin{center}
{\large \bf Discriminating Minimal SUGRA and Minimal
Gauge Mediation Models at the Early LHC
}
\vskip 1.2cm
Shoji Asai${}^1$, Eita Nakamura${}^1$ and Satoshi Shirai${}^{2,3}$
\vskip 0.4cm

{\it
$^1$Department of Physics, University of Tokyo, \\
Tokyo 113-0033, Japan\\
$^2$Department of Physics, University of California, \\Berkeley, CA 94720\\
$^3$ Theoretical Physics Group, Lawrence Berkeley National Laboratory, \\Berkeley, CA 94720
}

\vskip 1.5cm

\abstract{
Among various supersymmetric (SUSY) standard models,
the gravity mediation model with a neutralino LSP and the gauge mediation model with a very light gravitino
are attractive from the cosmological view point.
These models have different scales of SUSY breaking and their underlying physics in high energy is quite different.
However, if the sparticles' decay into the gravitino is prompt in the latter case, their collider signatures can be similar:
multiple jets and missing transverse momentum.
In this paper, we study the discrimination between these models in minimal cases at the LHC based on the method
using the significance variables in several different modes and show the discrimination is possible at a very early stage
after the discovery.
}
\end{center}
\end{titlepage}

\setcounter{page}{2}

\section{Introduction}

Supersymmetric (SUSY) standard model (SM) is the most promising candidate for the physics beyond the standard model (BSM).
Once its presence is inferred from experiments, the next problem is to search for the underlying mechanism
of SUSY breaking.
The scale of SUSY breaking is one of the most important clue for this search.

There are many proposed models with various SUSY breaking scales.
A major candidate is the gravity mediation model with a medium SUSY breaking scale.
In the model, the lightest SUSY particle (LSP) is the Bino-like neutralino.
It is stable and becomes a good candidate of a WIMP dark matter (DM),
if $R$-parity conservation assumed.
At the LHC, the LSP is observed as a missing transverse momentum ($p_T$).
The missing $p_T$ serves as a distinctive signature as well as high-$p_T$ jets and other decay products
coming from decay chains of produced SUSY particles.

Another attractive candidate is the gauge mediation (GM) model with a low SUSY breaking scale
since it  solves the SUSY flavor problem.
The gravitino becomes the LSP in the GM model
and the collider signatures strongly depend on the gravitino mass $m_{3/2}$,
which is related to the SUSY breaking scale $\sqrt{F}$ as $m_{3/2}=F/(\sqrt{3}M_{PL})$, and the nature of the next LSP (NLSP).
In the case of the heavier gravitino, the SUSY event at the LHC involves a non-pointing NLSP decay
\cite{hep-ph/0309031,arXiv:1111.3725} 
or a long-lived NLSP \cite{Fairbairn:2006gg},
which becomes a characteristic signature.
On the other hand, if the gravitino is very light,
the NLSP decay is prompt
and the gravitino is observed as a missing $p_T$.
Then, the event signature resembles that of the gravity mediation model.
From the cosmological point of view, such a very light gravitino with $m_{3/2}\lesssim10$ eV is very attractive since
the model is free from gravitino problems \cite{gravitinoprob}.
So the discrimination between the gravity mediation model and the GM model with a very light gravitino
is an important problem.
However, this discrimination is difficult.
In principle, measuring the LSP mass provides good information for discrimination.
But, to measure the LSP mass precisely, a large integrated luminosity is required, 
and in some case, obtained result may have degenerated solutions.

In this paper, we concentrate on the minimal types of the  gravity mediation model and the GM model.
We call these models the mSUGRA model and mGMSB model, respectively.
The NLSP in the mGMSB model becomes either the Bino-like neutralino $\tilde{\chi}^0_1$ or the lighter slepton  $\tilde{\ell} $ (stau $\tilde{\tau}_1$).
In the former case photons from the Bino decays are produced and in the latter case leptons are produced,
and the photons or leptons are expected to be the discriminant of the GM model.
High energy photons are rarely produced in the gravity mediation model and there is no difficulty in discriminating the GM
model with a Bino NLSP.
However, leptons are also produced in the gravity mediation models and thus the discrimination has a more quantitative nature.
Also, in the experiment, the signal events are smeared by the SM backgrounds and the effect of detector efficiencies.
Furthermore, if the lightest neutralino mass rests between the lighter slepton and stau: $m_{\tilde{\tau}_1}<m_{\tilde{\chi}^0_1}<m_{\tilde{\ell}_{R}}$, the lepton emissions can be
suppressed and/or hidden. 

The purpose of this paper is to study this discrimination between the mSUGRA and the mGMSB models in the realistic experimental setup.
We propose a simple method of discrimination and show the integrated-luminosity dependence of the degeneracy of models.
The method is applicable at an early stage of the LHC soon after the discovery.

The organization of the paper is as follows: In the next section, we describe the parameter spaces and the mass spectrums
of the models considered in this paper and introduce the collider signatures of these models.
We present the result of our Monte Carlo simulation and show the idea of using significance variables of multiple modes
in Section \ref{simulation}. 
In Section \ref{SIGTEST}, we
discuss the discrimination of the mGMSB and the mSUGRA models
based on a test on significance variables.
Section \ref{conclusion} is devoted for discussions and conclusions.

\section{mSUGRA vs mGMSB: event signatures}\label{signature}
In this section, we first review the parameter space of each model and
discuss the LHC signatures of these models.
Then we describe the idea of
discriminating these models at an early stage at the LHC.

\subsection{Model parameters and mass spectrum}\label{model}

In this paper, we compare the collider signatures of the mSUGRA model and the mGMSB model.
First, we describe the model parameters of each model.

The mSUGRA model represents a mass spectrum of the gravity-mediated SUSY breaking model
assuming the flavor blindness for the scalar masses and the CP property.
The model parameters are $m_0$, $m_{1/2}$, $A_0$, ${\rm tan}\,\beta$ and ${\rm sign}\,\mu$,
which is the unified scalar mass, the unified gaugino mass, the universal $A$-term, the ratio of the Higgs VEVs and
the sign of $\mu$, respectively. The low-energy mass spectrum is obtained by solving the renormalization-group equations
with these input values at the GUT scale. The resulting LSP is typically the lightest neutralino or the lightest slepton.
From the cosmological view point, we only consider the case of a neutralino LSP.

As a parameter space of the mGMSB model, we adopt a parametrization used in Ref.~\cite{Nakamura:2010faa}.
Here, the ratio between the gaugino mass and the sfermion mass
is left free while the ``GUT relations'' among the gaugino masses and the sfermion masses are imposed.
The $A$-term is set to be 0 and the $B$-term is tuned to give a correct low-energy electroweak parameters.
The model parameters are $\Lambda_s$, $\Lambda_g$, $M_{\rm mes}$, ${\rm tan}\,\beta$
and ${\rm sign}\,\mu$. The sfermion masses and the gaugino masses at the messenger mass scale $M_{\rm mes}$
are given by
\begin{align}
\label{eq:gaugino_mass2}
M_{a} \;&=\; \frac{\alpha_a}{4\pi}\Lambda_g,\\
\label{eq:scalar_mass2}
m^2_{\phi_i}\;&=\;\Lambda_s^2 \sum_a \left(\frac{\alpha_a}{4\pi}\right)^2 C_a(i),
\end{align}
where $\alpha_a$ ($a=1,2,3$) are the SM gauge couplings and $C_a(i)$ are the quadratic Casimir
coefficients for each field. The $\Lambda_s$ and $\Lambda_g$ are related to the parameters of
the conventional mGMSB model parameters as
\begin{align}
\Lambda_s^2&= 2N_5\Lambda^2,\\
\Lambda_g&= N_5\Lambda,
\end{align}
where $N_5$ is the number of messengers in the ${\bf 5}+{\bf5}^*$ representation and $\Lambda=F/M_{\rm mes}$
is the ratio of the square of the SUSY-breaking scale $\sqrt{F}$ and $M_{\rm mes}$,
when  $F/M^2_{\rm mes}  \ll 1$.
So in other words, the parameter representation in Eqs. (\ref{eq:gaugino_mass2}) 
and (\ref{eq:scalar_mass2}) is the same as the conventional one with an arbitrary (non-integer) number of messengers.

\subsection{mSUGRA signal}
The LSP is the lightest neutralino in the mSUGRA model as described in the previous section.
Once the SUSY particle, mostly the gluino and the squarks, is produced
it decays in a cascade down to the SM particles and the LSP.
Since the stability of the LSP is guaranteed by the $R$-parity conservation,
the produced LSPs then escape outside the detector and are observed as a missing $p_{\rm T}$.

Since the gluinos and the squarks are colored particles and the LSP is a color-neutral particle,
the cascade decay products include jets, as well as other SM particles depending on the
detailed form of the decay chain. 
There are many possible decay chains.
For example, in the decay chain $\tilde{q}_L \to \tilde{\chi}^2_0 q \to \tilde{\chi}^1_0 \ell \ell q$,
leptons are emitted as well as jets and missing $p_{\rm T}$.
Likewise, the decay products may include leptons, tau-jets or $b$-jets as well as the generic signal of jets and missing momentum.

\subsection{mGMSB signal}
In the mGMSB model we are interested in, the LSP is the gravitino.
As in the mSUGRA case, the produced SUSY particles decay into the gravitino LSP.
In the mGMSB case, however, the role of the NLSP becomes more important.

The decay length of the NLSP into the gravitino LSP is given by
\begin{equation}
c\tau\sim20\,\mu{\rm m}\left(\frac{m_{3/2}}{1~{\rm eV}}\right)^2\left(\frac{m_{\rm NLSP}}{100~{\rm GeV}}\right)^{-5}.
\end{equation}
Since this is much longer than the one resulting from the SM interaction, a produced SUSY particle usually decays
into the NLSP before it directly decays into the LSP.
\footnote{
There are exceptional cases. One example is the case when the right-handed slepton mass is degenerated to the lighter stau mass
for small ${\rm tan}\,\beta$ \cite{Hamaguchi:2007ge}.
Another example is the case that the masses of the lightest neutralino and the lighter stau are degenerated for 
$\L_g \sim \L_s$.
}
Thus, the event signature of the mGMSB model strongly depends on which particle is
the NLSP. 

As mentioned in the previous section, the NLSP in the mGMSB model is either the lightest neutralino or 
the lightest slepton (stau). In the neutralino-NLSP case the NLSP emits a photon  and in the slepton-NLSP
case the NLSP emits a lepton with a sizable branching fraction. The typical mGMSB signal, therefore, contains multi-jets $+$ missing $+$ multi-photons
or multi-jets $+$ missing $+$ multi-leptons.

The lightest slepton is the lighter stau $\tilde{\tau}_1$ and  the NLSP
in the slepton-NLSP case. So as well as the multi-leptons signal, we expect events with multi tau-jets.
The multi-tau-jets mode becomes more important in the large ${\rm tan}\,\beta$ case,
since then the heavier sparticles' decays into staus becomes more dominant because of the kinematical preference
and the enlarged coupling of the $\tilde{\tau}_1$ due to the large left-right mixing of the staus.

\subsection{Strategy for discrimination}
As described above, an important difference between the mSUGRA model and the mGMSB models arises
from the important role of the NLSP decay in the mGMSB model. In the mGMSB model with a neutralino-NLSP,
the signal event usually accompanies two photons, and in the case with a slepton-NLSP, the signal accompanies multi-leptons.
Grounded on these facts, we compare the models by using the following modes: no lepton $+$ 2 jets, no lepton $+$ 4 jets,
same-sign (SS) 2 leptons, 2 tau-jets and 2 photons modes.
\begin{table}[htbp]
\caption{Naively expected number of signal events.
The mark $\circledcirc$ indicates that a large number of signals are expected and
the mode is qualitatively ``good'' for model search. Similarly, $\bigcirc$, $\triangle$ and $\times$
indicates that the corresponding mode is qualitatively ``not bad'', ``not good'' and ``bad''.}
\label{mode_table}
\begin{center}\begin{tabular}{|c|c|c|c|}
\hline
mode & mSUGRA & mGMSB $\tilde{\tau}$-NLSP & mGMSB $\tilde{\chi}^0_1$-NLSP \\ \hline
$0\ell$ $+$ (2 jets/4 jets) &$\circledcirc$&$\triangle$&$\triangle$\\
SS $2\ell$ &$\bigcirc$&$\circledcirc$&$\triangle$\\
$2\tau$ &$\triangle$&$\circledcirc$&$\triangle$\\ 
$2\gamma$ &$\times$&$\times$&$\circledcirc$\\
\hline
\end{tabular}\end{center}
\end{table}
In table \ref{mode_table}, we summarize the naively expected number of signal events in these modes for each model.
For example, in the mSUGRA model, the no lepton $+$ (2 jets/4 jets) mode is expected to be
 the most viable and the modes of two leptons
or tau-jets are the next, but the two photons mode is bad because photons are hardly expected in this model.
Similarly, the two leptons mode or the two tau-jets mode is expected to be the most viable one for the mGMSB model
with a slepton-NLSP and the two photons mode is the best mode for the mGMSB model with a neutralino-NLSP.
Since the location of the viable mode differs for the three cases, we may distinguish these models by
observing which of the four modes shows the most signal events or which does not.

However, we need sophistication to apply this idea in the realistic experimental set-up. 
First, we must include contamination by the SM backgrounds and the detection efficiencies.
The signal events may be hidden unless we carefully reduce the SM background contribution, which might dilute the apparent
differences of characteristic events.
Further, there are a certain amount of misidentification and fake rate of particles by the detection systems.
This effect is non-negligible in the case of many leptons and photons, and is also a source of dilution of the signal events.
Second, we need a quantitative way to describe the distinctions among various models, which can incorporate
the experimental errors.

In the next section, we perform a Monte Carlo simulation, which incorporates the SM backgrounds and the detection efficiencies.
We use the significance variable (see Sec. \ref{sec_sig} for the definition and explanation)  to solve the above problems.

\section{mSUGRA vs mGMSB: simulation}\label{simulation}

In this section, we first describe the set-up of our Monte Carlo simulation and introduce the significance variable.
We also review the current experimental constraints.
We then show the result of our simulation and discuss the possibility of discrimination with significances in several different modes.

\subsection{SM backgrounds and detection efficiencies}

For the multi-jets mode, the main SM backgrounds come from the QCD jets, $t\bar{t}$ and gauge boson production events.
For the multi-lepton mode $t\bar{t}$ and $W+$ jets are the main background and for the multi-photon mode,
the $\gamma\gamma$  events are. To estimate these backgrounds and other sub-leading backgrounds, we
used the programs MC@NLO 3.42\cite{Frixione:2002ik}
(for $t\bar{t},WW,WZ$ and $ZZ$),
Alpgen 2.13\cite{Mangano:2002ea} (for $Wj,Zj$ and $W/Z+b\bar{b}/t\bar{t}$),
MadGraph 4.1.44 \cite{Alwall:2007st} (for $t\bar{t}/W/Z+\gamma/\gamma\gamma$), and
Pythia 6.4 \cite{Sjostrand:2006za} (for $\gamma\gamma$ and QCD jets).
As for the detector simulation, we used the AcerDet 1.0 \cite{RichterWas:2002ch}.
For the isolation criteria for leptons and photons,
we used the default setting of AcerDet.

In our simulation, we are interested in the signals with leptons or photons.
The detection efficiency and misidentification of these particles are important in estimating both the
signal events and background events.
In the following simulation, we include the misidentification and fake rates of the leptons, photons and jets.
The rates we have used in the analysis are shown in Table \ref{tb:fake}.
The values are taken from Ref. \cite{Aad:2009wy}.
For the SM backgrounds, however,
the efficiencies are chosen to give more conservative results than the actual experimental data.
\begin{table}[htbp]
\caption{Misidentification and fake rates. The efficiencies are taken from Ref. \cite{Aad:2009wy}.
For SM backgrounds, the efficiencies are chosen to give a conservative estimation.}
\label{tb:fake}
\begin{center}\begin{tabular}{|c|c|c|c|c|c|c|c|c|c|}
\hline
&$j\to e$ & $j\to \mu$&  $j \to \gamma$ &$\tau$ $\to j$  & $j\to\tau$& $e \to j$ &$\mu \to j$  & $e \to \gamma$   & $\gamma \to j$ \\ \hline
SUSY&0.3 \%  &0.3 \%  &0.02 \% & 60 \%&1 \% & 27 \% & 30 \% &3 \% & 20 \% \\ \hline
SM BKG&0.3 \% &0.3 \% &0.02 \% &20 \% & 1 \% &0 \% & 0 \% & 3 \%& 0 \% \\
\hline
\end{tabular}\end{center}
\end{table}

\subsection{Significance}\label{sec_sig}

Under a fixed set of event cuts, we estimate the event number of the  signals $N_s$ and the backgrounds $N_b$.
To compare many model points in several modes on the same footing, we use the significance variable $Z$,
which is a function of $N_s$ and $N_b$. We use the significance incorporating both the statistical and
systematic uncertainties of the backgrounds and adopt the one used in Ref. \cite{Aad:2009wy}. The
systematic uncertainties of the backgrounds are taken to be 50\% for the QCD multi-jet
background and 20\% for others.

Given the number of the signal events $N_s$ and the background events $N_b$ with the uncertainty $\delta N_b$,
the significance is given by calculating the convolution of the Poisson distribution with some ``posterior'' distribution
function. As the posterior distribution, we take the gamma distribution as suggested in Ref. \cite{Linnemann}.
The resulting significance $Z$ is given by \cite{Linnemann}
\begin{equation}
Z=\sqrt{2}\,{\rm erf}^{-1}(1-2p),
\end{equation}
with
\begin{equation}\label{prob}
p=\frac{B(N_s+N_b,1+N_b^2/\delta N_b^2,\delta N_b^2/(N_b+\delta N_b^2))}{B(N_s+N_b,1+N_b^2/\delta N_b^2)},
\end{equation}
where ${\rm erf}^{-1}$ is the inverse error function and
\begin{equation}
B(a,b,x)=\int_0^x dt\, t^{a-1}(1-t)^{b-1}
\end{equation}
is the incomplete beta function and $B(a,b)=B(a,b,1)$ is the usual beta function. 
If we take the limit $\delta N_b\to0$, the Eq.~(\ref{prob}) reduces to the probability
in the Poisson distribution.
The statistical error of the significance $\delta Z_B$ is depends on both $N_s$ and $N_b$.
The error is ${\cal O}(1)$ for the moderate values which we are interested in.

\subsection{Experimental constraints}\label{constr}
Up to now, the most stringent experimental constraints are given by the ATLAS and CMS experiments at integrated luminosities
about $1\,{\rm fb}^{-1}$.
Analyses relevant for the GMSB and mSUGRA models are searches of jets with no lepton \cite{ATL_0ljets} or one lepton \cite{ATL_1ljets}
by the ATLAS collaboration and searches of  same-sign \cite{CMS_SS2l} or opposite-sign \cite{CMS_OS2l} dilepton 
and photons \cite{CMS_2g} by the CMS collaboration.

\begin{figure}
\subfigure[mGMSB ${\rm tan}\,\beta=10$.]{
\includegraphics[clip, width=0.48\columnwidth]{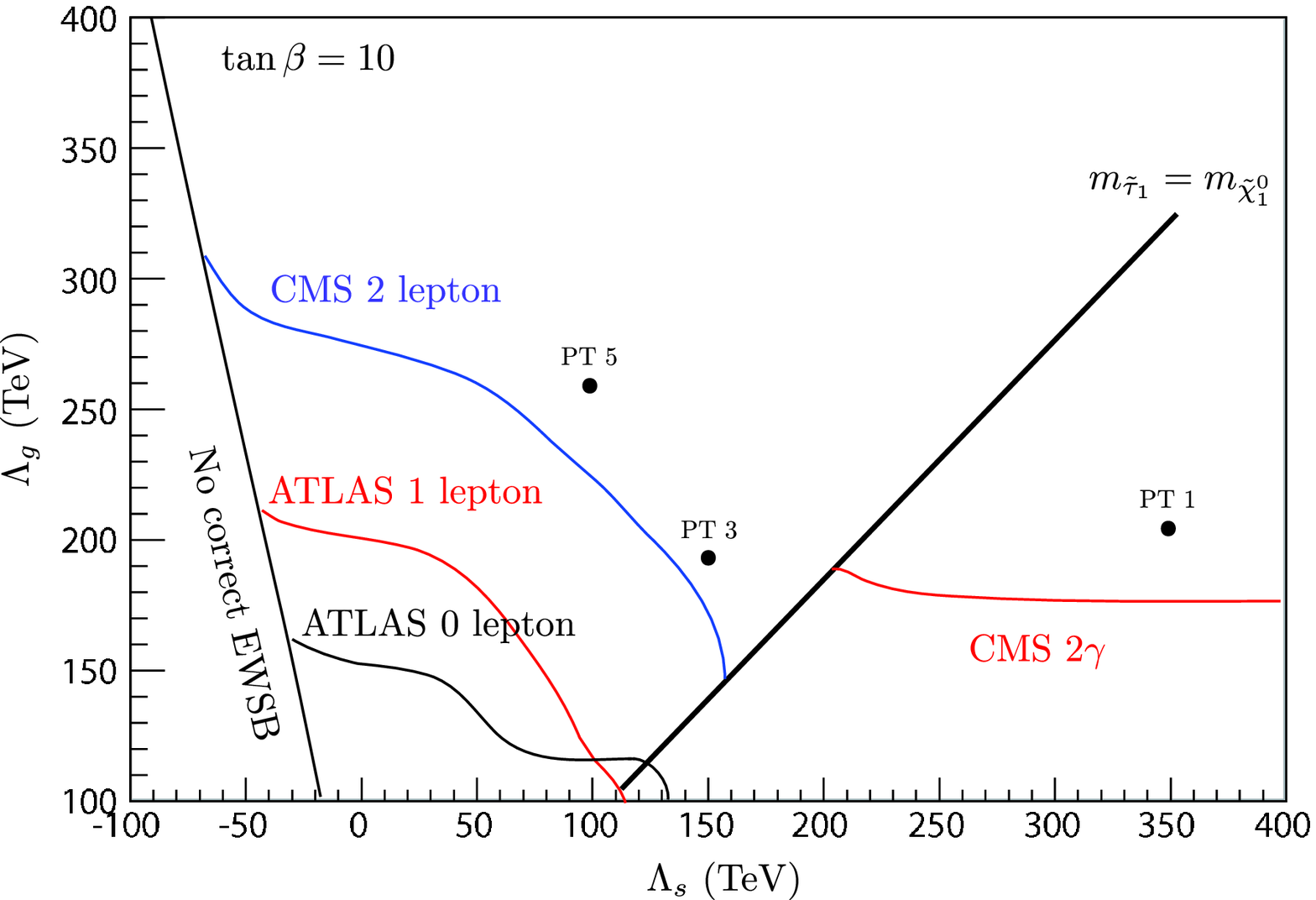}
}
\subfigure[mGMSB ${\rm tan}\,\beta=40$.]{
\includegraphics[clip, width=0.48\columnwidth]{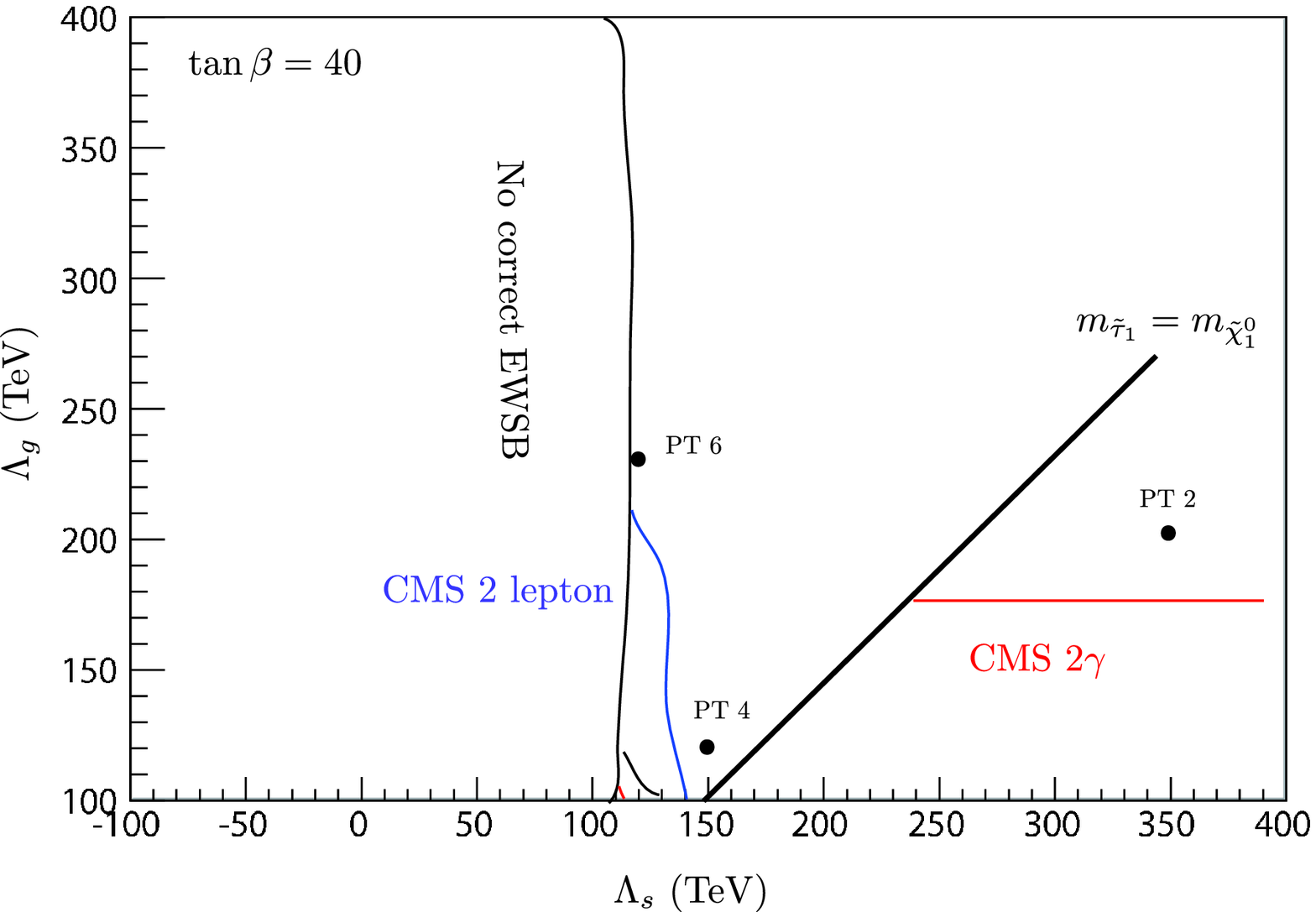}
}\\
\subfigure[mSUGRA ${\rm tan}\,\beta=10$, $A_0=0$.]{
\includegraphics[clip, width=0.48\columnwidth]{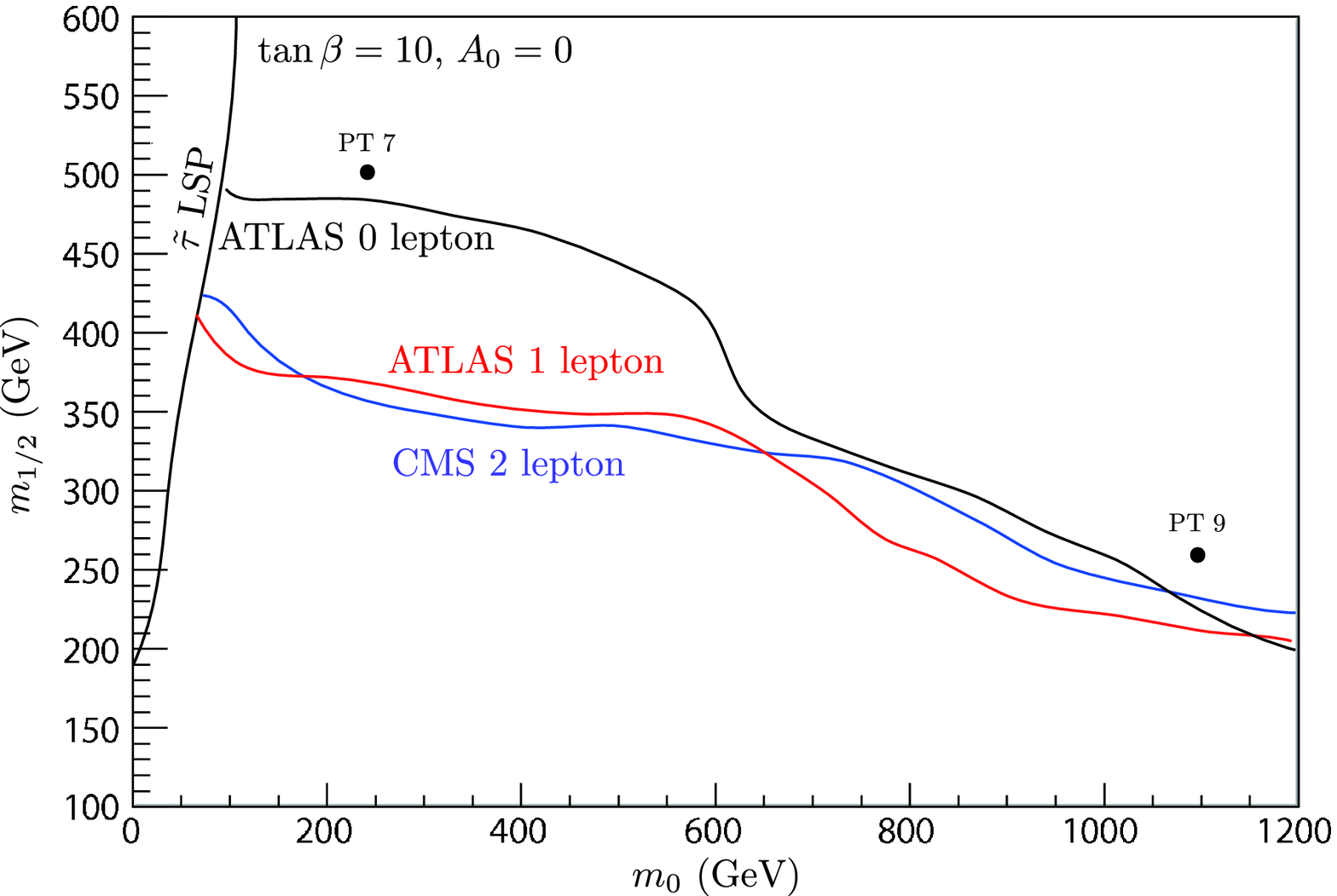}
}
\subfigure[mSUGRA ${\rm tan}\,\beta=40$, $A_0=0$.]{
\includegraphics[clip, width=0.48\columnwidth]{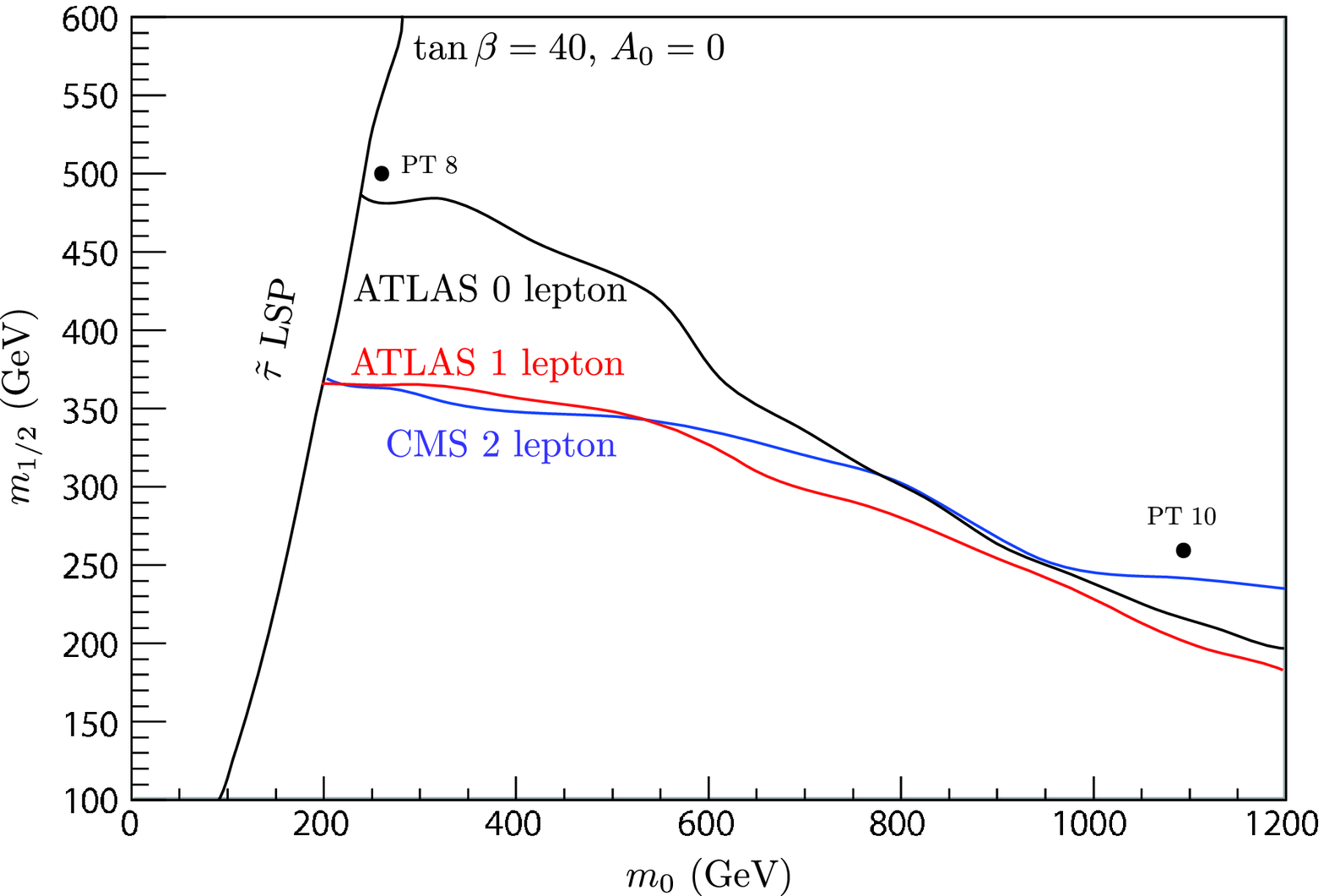}
}
\caption{Experimental constraints given by the LHC experiment at an integrated luminosity of about 1 ${\rm fb}^{-1}$
at the 7 TeV run.
We also show the benchmark points PT1--10.
}
\label{fig_constr}
\end{figure}

We show the constraints given by these experimental data in Fig. \ref{fig_constr}.
To obtain the constraints, we apply the 95\% C.L. exclusion limits given by the above analyses to our model points.
To incorporate the NLO corrections to the cross section, we multiply a uniform $k$-factor of 1.5 for all model points,
except for the mGMSB model points with a Bino-NLSP, at which the dominant contribution to the SUSY production
is given by the direct Wino production.

\subsection{Signal events}\label{sec_signal_event}
For both the mSUGRA and mGMSB models,
the MSSM mass spectrums and the decay tables are calculated by using the program
ISAJET 7.72 \cite{ISAJET}.
The SUSY events are generated by using the Herwig 6.510 \cite{HERWIG6510}
and we have used the AcerDet 1.0 \cite{RichterWas:2002ch} for the detector simulation.

For the mSUGRA model, the grid in parameter space is taken to be $0\leq m_0\leq 1200$ GeV (in steps of 20 GeV),
$100\leq m_{1/2}\leq600$ GeV (in steps of 20 GeV), $A_0=0,$ $m_0$, $-m_0$, ${\rm tan}\,\beta=10,$ 40
and ${\rm sign}\,\mu=+$.  
For the mGMSB model, the grid in parameter space is
$-(100)^2\leq\Lambda_s^2\leq(400)^2$ TeV$^2$ (in steps of 5 TeV), $(100)^2\leq \Lambda_g^2\leq(400)^2$ TeV$^2$ (in steps of 5 TeV) and
${\rm tan}\,\beta=10$, 40. The messenger mass scale is fixed as $m_{\rm mes}=200$ TeV \footnote{
To avoid a messenger tachyon,  the condition $F \lesssim m_{\rm mess}^2$ is required, which indicates $m_{\rm mess} \sim 100$ TeV.
}
 and 
the gravitino mass is set $m_{3/2}=16$ eV. 

Because we are interested in the model points that are potentially discovered at an early stage of the LHC and are not excluded by the experiments,
we select model points within a certain mass region to reduce the size of computational calculation. 
For the mSUGRA, we select model points with $800<m_{\tilde{d}_R}<1600$ GeV and $400<m_{\tilde{g}}<1400$ GeV.
For the mGMSB,
we select model points with $850<m_{\tilde{d}_R}<1600$ GeV for the slepton-NLSP case and $220<m_{\tilde{\chi}^0_1}<350$ GeV for the neutralino-NLSP case.
The number of model points after this selection is 5633 (4046) for the mSUGRA (mGMSB) model.
After imposing the experimental constraints reviewed in the previous section the number reduces to 3654 (3048).

\subsection{Modes and kinematical cuts}
In the following analysis, we use five different modes, no lepton modes with two and four jets, two photons mode, two $\tau$-jets mode,
and the same-sign (SS) two leptons mode.

The kinematical cuts for these search modes are summarized in Table \ref{cuts}.
\begin{table}[htbp]
\caption{List of kinematical cuts used in the analysis.
The $E_{\rm T}^{\rm miss}$ is the absolute value of the missing $p_{\rm T}$ and
$M_{\rm eff}$ is defined in Eq. (\ref{eq_meff}).
We require isolation of leptons and photons (see text).
In the four jets with no lepton mode, the values in brackets are chosen for each model point
to optimize the significance.}
\label{cuts}
\begin{center}\begin{tabular}{|c|l|}
\hline
mode&\\\hline
SS $2\ell$	& At least two leptons with $p_{\rm T}>20$ GeV. The leading two leptons have \\
			& the same charge. At least 2 jets with $p_{\rm T}>50$ GeV and for the leading\\
			& jet $p_{\rm T}>100$ GeV. $E_{\rm T}^{\rm miss}>100$ GeV, $M_{\rm eff}>800$ GeV. \\\hline
$2\tau$		& At least two tau-jets with $p_{\rm T}>40$ GeV. At least 4 jets with \\
			& $p_{\rm T}>50$ GeV and for the leading jet $p_{\rm T}>100$ GeV. \\
			& $E_{\rm T}^{\rm miss}>100$ GeV, $M_{\rm eff}>1000$ GeV. \\\hline
$2\gamma$		& At least two photons with $p_{\rm T}>60$ GeV and for the leading photon \\
			& $p_{\rm T}>90$ GeV. $E_{\rm T}^{\rm miss}>100$ GeV\\\hline
$0\ell+4$ jets	& At least four jets, requiring $p_{\rm T}>\{50,100\}$ GeV, and for the leading jet\\
			& $p_{\rm T}>100$ GeV. No leptons with $p_{\rm T}>10$ GeV.  $E_{\rm T}^{\rm miss}>200$ GeV,\\
			& $M_{\rm eff}>\{1000,1200\}$ GeV and $E_{\rm T}^{\rm miss}/M_{\rm eff}>\{0.2,0.3\}$. \\\hline
$0\ell+2$ jets	& At least two jets with $p_{\rm T}>200$ GeV, and for the leading jet\\
			& $p_{\rm T}>300$ GeV. No leptons with $p_{\rm T}>10$ GeV.  $E_{\rm T}^{\rm miss}>400$ GeV,\\
			& $M_{\rm eff}>1200$ GeV and $E_{\rm T}^{\rm miss}/M_{\rm eff}>0.35$. \\
\hline
\end{tabular}\end{center}
\end{table}
In the table, the $E_{\rm T}^{\rm miss}$ denotes the absolute value of the missing $p_{\rm T}$, and
the effective mass is $M_{\rm eff}$ defined by
\begin{equation}\label{eq_meff}
M_{\rm eff}=E_{\rm T}^{\rm miss}+\sum_{j:\text{ jet}}^{\text{2 or 4}}p_{\rm T}(j)+\sum_{\ell:\text{ lepton}}p_{\rm T}(\ell),
\end{equation}
where the summation on the jet $p_{\rm T}$ is for the leading two jets in the two $\tau$-jets mode and the no lepton with two jets mode,
and for the leading four jets in other modes.
The cuts in each mode are those typically used by the current experiment at the LHC.
To fix the numerical values for kinematical variables, we performed simple optimization to maximize the significance.
In the four jets with no lepton mode, for which the maximized values depend rather heavily on the location of the model points,
we left several cuts unfixed.
When a cut has two values in brackets, it means that the kinematical cut is optimized within the described values.

\subsection{Example model points}
We first pick up several example model points to illustrate the result.
The 10 selected model points (PT 1 to PT 10) are depicted in Fig.~\ref{fig_constr} and the corresponding
model parameters are summarized in Table \ref{ex_pts}.

\begin{table}[htbp]
\caption{Model parameters of the example model points.}
\label{ex_pts}
\begin{center}\begin{tabular}{|c|c|c|c|c|}
\hline
mGMSB & $\Lambda_s$ (TeV) & $\Lambda_g$ (TeV) & ${\rm tan}\,\beta$ & \\ \hline
PT 1 & 350 & 200 &10 & \\ \hline
PT 2 & 350 & 200 &40 & \\ \hline
PT 3 & 150 & 180 &10 & \\ \hline
PT 4 & 150 & 120 &40 & \\ \hline
PT 5 & 100 & 260 &10 & \\ \hline
PT 6 & 120 & 230 &40 & \\ \hline
mSUGRA & $m_0$ (GeV) & $m_{1/2}$ (GeV) & ${\rm tan}\,\beta$ & $A_0$ \\ \hline
PT 7 & 260 & 500 &10 & 0 \\ \hline
PT 8 & 260 & 500 &40 & 0 \\ \hline
PT 9 & 1100 & 260 &10 & 0 \\ \hline
PT 10 & 1100 & 260 &40 & 0 \\ \hline
\end{tabular}\end{center}
\end{table}
\begin{figure}
\begin{center}
\includegraphics[clip, width=0.6\columnwidth]{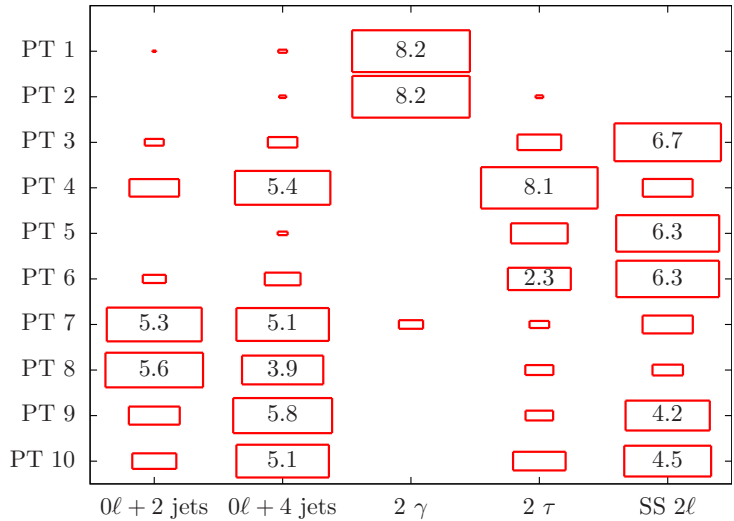}
\caption{Significances $Z$'s of the example model points for the five modes listed in Tab. \ref{cuts}.
The size of red boxes shows the corresponding significance with an integrated luminosity of 30 ${\rm fb}^{-1}$ at 7 TeV run.}
\label{fig_siglist}
\end{center}
\end{figure}

PT 1 to 6 are mGMSB model points.
PT 1 and 2 represent models with a Bino-NLSP,
PT 3 and 4 models with a slepton-NLSP but the $\tilde{\tau}_1$ mass and the lightest neutralino mass
are nearly degenerate, and PT 5 and 6 with a slepton-NLSP without such a degeneracy.
PT 7 to 10 are mSUGRA model points. Here, we take $A_0=0$.
PT 7 and 8 are typical models with small $m_0$ and large $m_{1/2}$,
and PT 9 and 10 are typical models with large $m_0$ and small $m_{1/2}$.
In the Appendix, we list the mass spectrum of relevant superparticles and their
dominant decay modes in Tab. \ref{tab_mass} and \ref{tab_mass_2}.

In Fig.~\ref{fig_siglist}, we illustrate the significances of the example model points in the five mode
in Tab. \ref{cuts}
with the integrated luminosity of 30 ${\rm fb}^{-1}$ at 7 TeV run.
The number of signal events and our estimation of the SM background
is listed in Tab. \ref{NsNb} in the Appendix.
We see that the naive expectation in Table \ref{mode_table} is in fact
visible in terms of the significances.
An exception is PT 4, where the no lepton modes have unexpectedly large significances.
At this point, the lighter stau mass is nearly degenerated to the lightest neutralino mass and
the significant proportion of taus or leptons from the upper-stream  of decay-chains become soft.
This fact results in smaller events with high-$p_{\rm T}$ leptons.

\subsection{Results for all model points}\label{MC}
We now illustrate results for all model points.
Since the significance variable is an increasing function of the number of signal events after the cut,
its absolute value scales with the total cross section of the model point.
Thus the meaningful quantity for discrimination is the ratio of significances in different modes or a similar quantity.
In the following we show the distribution of the significance in each mode in comparison with the one in the four jets with no lepton mode $Z(0\ell4j)$.
The significances are calculated for the integrated luminosity of 30 ${\rm fb}^{-1}$ at 7 TeV run.

\begin{figure}
\begin{center}
\subfigure[$Z(2\gamma)$ vs. $Z(0\ell4j)$.]{
\includegraphics[clip, width=0.48\columnwidth]{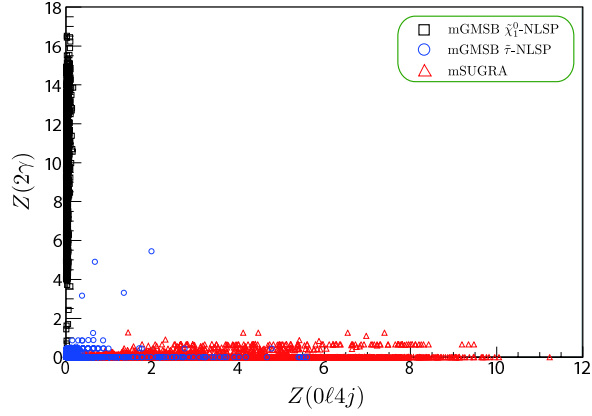}\label{fig_Z2pg}
}\end{center}

\subfigure[$Z({\rm SS}2\ell)$ vs. $Z(0\ell4j)$.]{
\includegraphics[clip, width=0.48\columnwidth]{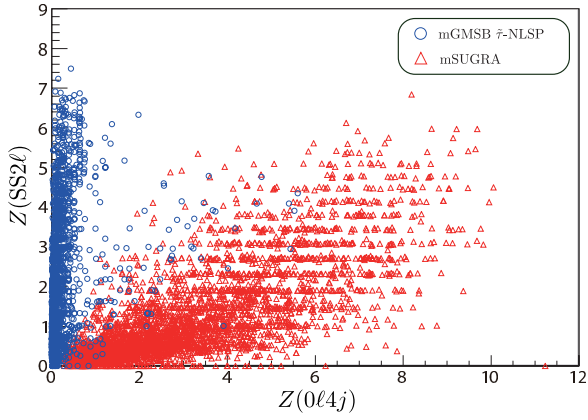}\label{fig_ZSS2pl}
}
\subfigure[$Z(2\tau)$ vs. $Z(0\ell4j)$.]{
\includegraphics[clip, width=0.48\columnwidth]{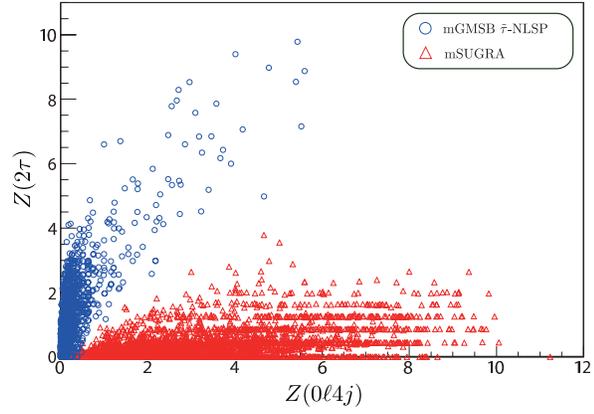}\label{fig_Z2ptau}
}

\subfigure[$Z(0\ell2j)$ vs. $Z(0\ell4j)$.]{
\includegraphics[clip, width=0.48\columnwidth]{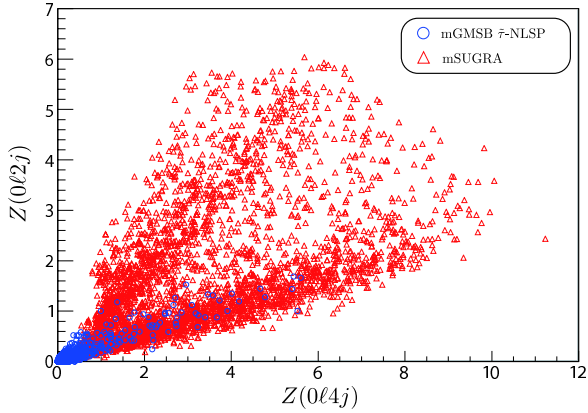}
\label{fig_Z0l2j}
}
\subfigure[$Z({\rm SS}2\ell)+Z(2\tau)$ vs. $Z(0\ell4j)+Z(0\ell2j)$.]{
\includegraphics[clip, width=0.448\columnwidth]{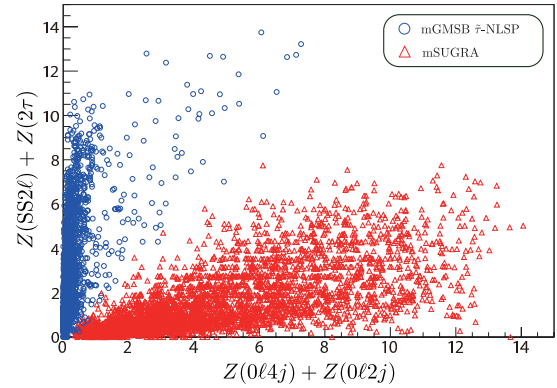}
\label{fig_ZjZl}
}
\caption{Scatter plots of significances  $Z$'s for an integrated luminosity of 30 ${\rm fb}^{-1}$ at 7 TeV run.
The black square points are the mGMSB model points with a neutralino-NLSP,
the blue circle points the mGMSB model points with a slepton-NLSP,
the red triangle points the mSUGRA model points.}\label{s_sig}
\end{figure}

We first compare the photon mode significance.
Since the multi-photon signal is a very distinctive signature for the mGMSB model with a neutralino-NLSP, 
the model is expected to be distinguished clearly from the others in terms of the photon mode significance.
In Fig. \ref{fig_Z2pg}, we plot the photon mode significance $Z(2\gamma)$ versus the four jets mode significance $Z(0\ell4j)$.
Each point in the figure corresponds to a model point.
A red and triangle (blue and circle, black and square) mark represents a mSUGRA (mGMSB with a slepton-NLSP, mGMSB with a neutralino-NLSP) model point.
We can check a clear separation of the  mGMSB model points with a neutralino-NLSP from the other models as expected.
A few model points of the GMSB model with a slepton-NLSP have the significance larger than five.
At these model points, the lightest neutralino $\tilde{\chi}^0_1$ and the lighter slepton are almost degenerated and
the decay of the lightest neutralino $\tilde{\chi}^0_1$ into the lighter slepton is kinematically suppressed.
Thus, these model points are parametrically very near to the neutralino-NLSP case and we here treat them as
the neutralino-NLSP case. 

Next we show the significance in the SS two leptons mode $Z({\rm SS}2\ell)$. 
To clarify the illustration, we deselect the mGMSB model points with a neutralino-NLSP (or more precisely those with $Z(2\gamma)>3$). 
In Fig. \ref{fig_ZSS2pl} (and the following figures), the rest mGMSB model points are marked with blue and circle and the mSUGRA model points are marked with red and triangle. 
We see a clear inclination that the mSUGRA model points have smaller $Z({\rm SS}2\ell)/Z(0\ell4j)$ and
the mGMSB model points have larger $Z({\rm SS}2\ell)/Z(0\ell4j)$, showing that for most of the model points the lepton richness is indeed a good discriminator
even with the smearing due to the SM background and detector effects.
However, we also find many mGMSB model points with rather small  $Z({\rm SS}2\ell)/Z(0\ell4j)$.
As discussed in the previous section, a typical reason for a smaller $Z({\rm SS}2\ell)$ is nearly degenerated NLSPs. 
If the slepton-NLSP is nearly degenerated to the lightest neutralino, then a possible leptons coming from the decay of the lightest neutralino into the slepton becomes
soft and difficult to be detected. Another but related cause for a less lepton signature is larger ${\rm tan}\,\beta$.
As indicated in the previous section, when the ${\rm tan}\,\beta$ is large, the SUSY decay chains involve more staus than other sleptons,
because large $\tan \beta$ induces the large left-right mixing of the staus, which
leads to the small stau mass and large interaction originated from the component of the left-handed stau.
Thus decay into the stau is preferred from both reasons of kinematics and interaction strength.
Note that the decrease of the leptons results in not only a smaller $Z({\rm SS}2\ell)$ but also a larger $Z(0\ell4j)$,
showing that a smaller $Z({\rm SS}2\ell)$ at those points is not cased by a mere scale of cross section.

To rescue the mGMSB model points with a small lepton mode significance,
we now show the significance in the two $\tau$-jets mode versus that in the four jets mode in Fig. \ref{fig_Z2ptau}.
Here we see an apparently better separation between the mSUGRA and the mGMSB models compared to the dilepton mode.
Especially, those mGMSB model points with larger $Z(0\ell4j)$ have larger $Z(2\tau)$.
This is another check that in those model points the leptons are replaced by $\tau$-jets. 
For the discrimination, this $\tau$-jets mode is important
since it plays a complementary role to the dilepton mode.
Note that for most of the mGMSB model points, the $Z(2\tau)$ is small compared to $Z({\rm SS}2\ell)$.
This is because of the low detection efficiency of the $\tau$-jet.

Finally we look at the two jets mode in Fig. \ref{fig_Z0l2j}.
The mSUGRA model points are spread over a wide region, but the mGMSB model points have smaller $Z(0\ell2j)/Z(0\ell4j)$.
The two jets mode alone does not work as a discriminator, but it strengthen the separation between the mGMSB model points
and part of the mSUGRA model points.
In summary, with the inclusion of the SM background and detection efficiencies, the photon mode, SS dilepton mode together with two $\tau$-jets mode
and their counters, two jets and four jets with no lepton modes, remain serving as discriminator for the mSUGRA and mGMSB models.
In Fig. \ref{fig_ZjZl} we plot two sums $Z({\rm SS}2\ell)+Z(2\tau)$ vs. $Z(0\ell4j)+Z(0\ell2j)$ for both models.
We see a fairly good separation between the two models.

In reality, the significance values have uncertainties due to the experimental errors and errors from the MC statistics.
With an increasing integrated luminosity, each model point gains larger absolute values of the significances and thus the distance from the opposite set of models
(mSUGRA vs. mGMSB) become larger.
Then the next problem is how much the integrated luminosity is needed in order that 
the model point is well-separated from the opposite models.
In the next section, we study the integrated-luminosity dependence of the separation between the two model.

\section{Significance test for discriminating models} \label{SIGTEST}

Given the above results, we now go on to the discrimination of the mGMSB and the mSUGRA models.
Discrimination here means that excluding the mGMSB models from the mSUGRA models and vice versa.
In the previous section, we have seen that even with the inclusion of the effects of the SM background events
and the statistical uncertainty, we expect a fairly good separation between the two models at earlier stages
of the experiment after the discovery.
However, to answer the question of how far and at which luminosity the discrimination is possible
we need a further machinery.
In this section, we develop a statistical method and answer this question.

The statistical method we use is based on the $\chi^2$ test on the significance variables.
Given two model points, $m_a$ and $m_b$, and an integrated luminosity $L$, the $\chi^2$ variable is defined as
\begin{equation}\label{chi2}
\chi^2(m_a,m_b;L)=\sum_{i:{\rm modes}}\frac{\left(Z(m_a,i;L)-Z(m_b,i;L)\right)^2}{\left(\delta Z(m_a,i;L)\right)^2+\left(\delta Z(m_b,i;L)\right)^2},
\end{equation}
where $i$ runs for modes used in the analysis and $\delta Z$ is the error of $Z$.
In general, $\delta Z$ contains various systematic errors as well as the statistical error.
In the following analysis, we only consider the expected experimental statistical error 
and the uncertainty from MC statistics.
We may also define $\chi^2$ by using functions of significance variables. Then the denominator of Eq.~(\ref{chi2}) is
replaced by the error of corresponding functions.

If, for the two model points, the calculated $\chi^2$ is smaller than a certain value, their difference of significances are considered as
statistically insignificant. In the following analysis, we use a $p$-value of 95\% for this criterion. The corresponding $\chi^2$ value
depends on the number of degrees of freedom, i.e. the number of significances used for the calculation of $\chi^2$.
We say that two model points are in the vicinity of each other if their $\chi^2$ is smaller than than this value.

Now, we consider the problem of testing whether a certain model $m_0$ is consistent as an mGMSB model or an mSUGRA model.
We first prepare a set of model points in these models. We denote by $M$ the set of model points and indicate as $M=G$ the set
of mGMSB model points and $M=S$ that of the mSUGRA models, respectively.
Then let us perform a $\chi^2$ test and collect those model points of $M$ in the vicinity of $m_0$. We denote this set of model points as $S_M(m_0)$.
If $S_M(m_0)$ is not empty, there is a certain probability that $m_0$ is in the model of $M$.
However, we must take account of the density of points in $M$. If there are a large number of points in $M$,
there might be a high probability of the $S_M(m_0)$ being non-empty even if $m_0$ is not a good candidate in the model of $M$.
On the other hand, if the cardinality of $M$ is small there may be a possibility that $S_M(m_0)$ is empty even if $m_0$ is actually in the model of $M$.

For this purpose, we first consider a point $m$ in $M$ as $m_0$.
For the care of this last problem, we must prepare a dense enough set $M$.
Then for each point $m$ in $M$, $S_M(m)$ contains at least a few points.
The number of points in $S_M(m)$ indicates the sensitivity of the significances in the vicinity of $m$.
We now consider a quantity $s_M(m)$ defined as
\begin{equation}\label{sM}
s_M(m)=\sum_{m'\in S_M(m)}\frac{1}{\# S_M(m')}.
\end{equation}
If the variation of $\# S_M(m)$ is smooth compared to the density of model points around $m$
in the region of $S_M(m)$, the value of $s_M(m)$ is near unity.
Similarly, we define for a general model point $m_0$
\begin{equation}\label{gsM}
s_M(m_0)=\sum_{m'\in S_M(m_0)}\frac{1}{\# S_M(m')}.
\end{equation}
If the density of $M$ is high enough and the variation of $\# S_M(m)$ is smooth,
we expect that $s_M(m_0)$ is also near unity if $m_0$ is indeed in the model of $M$.
Note that this condition is stronger than the mere condition that $S_M(m_0)$ is not empty.
In this method of discrimination, the condition for $m_0$ to be consistent with being in the model of $M$
is that $s_M(m_0)$ is near to unity within the errors, which are discussed further below.

Before discussing the errors, we have several remarks on this method.
Let us consider the limit of infinite density of $M$ and infinite luminosity.
Then the distribution of model points in $M$ in the significance space spreads over a certain region
and then $s_M$ for a given model is unity or zero, indicating the model is in or out of this region.
In other words, this method is a sort of in-or-out test and does not take the density distribution of $M$
in this limit. So this method is useful to discriminate well separated models, but may be useless for those models,
in which a tiny amount of model points spreads over a wide region.
In practice, we have a finite density of $M$ and a finite luminosity, and the region is approximated by discrete points.
In this case, the sufficiency of the density of $M$ may be tested by $S_M(m)$ for $m$ in $M$ as noted above.
There may be some points with a small $S_M(m)$ compared to the rest.
One reason for such a case is the existence of points, called ``cliffs'' in Ref. \cite{LHCinv}.
Around these points the variation of the significance is large compared to that of the parameter.
The boundaries in the parameter space or in the mass spectrum space may be a cause for such a case.
This case is often physically interesting and may require further and detailed studies.

The errors come from the effects of both the finite density and luminosity.
In a case of finite luminosity, we have a certain variation of the value of the $s_M(m_0)$
other than unity or zero if $m_0$ is near the boundary of $M$ in the significance space.
Assuming $M$ is uniformly distributed, the $s_M(m_0)$ at the boundary is equal to ${\rm log}\,2\simeq0.69$ (independent
of the dimension of the significances) and increases as $m_0$ moves from the boundary into the region of $M$ to a maximal
of $\sim1.2$ depending on the dimension. Away from the boundary $s_M(m_0)$ decreases to zero.
These values may vary well according to the shape and distribution of concerning model $M$.
Since we do not know the distribution of $M$ and the position of $m_0$ in the significance space a priori,
we handle this variation as an uncertainty and we take the value $\delta_{l} s_M(m_0)=0.5$ for this uncertainty
in the following analysis.
There are also errors on $s_M(m_0)$ coming from the finite density of $M$.
We adopt a rough approximation of this error by assuming that $S_M(m_0)$ shapes like a multi-dimensional
sphere in the space of significances. In this approximation, the error of $s_M(m_0)$ is given by
\begin{equation}
\delta_{g} s_M(m_0)=s_M(m_0)\,d\left(\frac{C_d}{\# S_M(m_0)}\right)^{1/d},
\end{equation}
where $d$ is the degrees of freedom of $\chi^2$ and $C_d=\pi^{d/2}/\Gamma(\frac{d}{2}+1)$.
As a result, the condition for the consistency of $m_0$ in the model $M$ can be written as
\begin{equation}\label{disc_cri}
|s_M(m_0)-1|<\delta s_M(m_0)=\sqrt{\delta_g s_M(m_0)^2+\delta_ls_M(m_0)^2}.
\end{equation}
We remark that this estimation of $\delta s_M$ is very rough and we assume that the variation
of $S_M$ in the parameter space is smooth.

We examine this method of model discrimination with the result of the Monte Carlo simulation in Sec. \ref{MC}.
We use as the set $M=G$ or $S$ the set of all simulated model points described in Sec. \ref{sec_signal_event}.
As test model points, we choose from the same set model points which pass the experimental constraints
and satisfy the discovery condition, for which we require at least one of the significances is greater than five for 30 ${\rm fb}^{-1}$.
The set of test model points are denoted by $G'$ and $S'$.
The number of model points are 4046, 1091, 5633 and 754 for $G$, $G'$, $S$ and $S'$, respectively.
We show the results for $\chi^2$ calculated in three different combinations of the modes listed in Table \ref{cuts}.
The first one uses the 0 lepton 4 jets mode, 2 photons mode and SS 2 leptons mode,
the second one uses the sum of two 0 lepton modes, 2 photons mode and the sum of 2 $\tau$-jets and 2 leptons modes,
and the third one uses all the five modes.
We call these three cases as T1 to T3.
The results are illustrated with integrated luminosities $L=1$, 3, 10, 30 and 100 ${\rm fb}^{-1}$ in the 7 TeV run.

We first check model points with small $\# S_M$.
In Table \ref{smallSM}, the number of model points in $M=G$ and $S$ with $\#S_M$ less than ten is presented.
\begin{table}[htbp]
\caption{The number of model points with $\#S_M<10$, which pass the experimental constraints.
The total number of model points are 3048 for $G$ and 3654 for $S$.
The numbers for $L=1$ and 3 ${\rm fb}^{-1}$ are all zero.}
\label{smallSM}
\begin{center}\begin{tabular}{|c|c|c|c|}
\hline
$L$ &  10 fb$^{-1}$ & 30 fb$^{-1}$ & 100 fb$^{-1}$ \\\hline\hline
$G$ in T1  & 0 & 1 & 2 \\\hline
$G$ in T2  & 1 & 3 & 3 \\\hline
$G$ in T3  & 0 & 2 & 4 \\\hline\hline
$S$ in T1  & 0 & 0 & 0 \\\hline
$S$ in T2  & 0 & 0 & 0 \\\hline
$S$ in T3  & 0 & 0 & 0 \\\hline
\end{tabular}\end{center}
\end{table}
Because most of the points with $\#S_M<10$ are near the artificial boundary in the model parameter space,
especially in the low mass region, and they are unlikely to affect the result of the test,
only those which pass the experimental constraints are counted in the table.
The few mGMSB model points in the table are those nearly on the boundary where the NLSP changes.
The vicinity of these points is a good example of a ``cliff''.
The followed result shows the effect of this region for discriminating the models,
but a further study might be important if the experiment reveals its possibility.

Now we present our result of the tests T1 to T3.
In Table \ref{sGsS}, $s_{G}$ and $s_S$ averaged in $G'$ or $S'$ are presented.
\begin{table}[htbp]
\caption{Evolution of the averaged $s_{G,S}$ with respect to the integrated luminosity $L$
for significance test T1 to T3.}
\label{sGsS}
\begin{center}\begin{tabular}{|c|c|c|c|c|c|c|}
\hline
&$L$ & 1 fb$^{-1}$ & 3 fb$^{-1}$ & 10 fb$^{-1}$ & 30 fb$^{-1}$ & 100 fb$^{-1}$ \\\hline\hline
T1&$s_G$ averaged in $G'$& 1.036&	1.042&	1.011&	1.009&	0.99\\\hline
&$s_S$ averaged in $G'$ &0.8768&	0.5584&	0.1107&	0.0412&	0.034\\\hline
&$s_G$ averaged in $S'$ &1.119&	0.7287&	0.4493&	0.4751&	0.4448\\\hline
&$s_S$ averaged in $S'$ &1.062&	1.202&	1.009&	0.9293&	0.9002\\\hline\hline
T2&$s_G$ averaged in $G'$ &1.031&	1.033&	1.01&	1.019&	0.997\\\hline
&$s_S$ averaged in $G'$ &0.823&	0.5146&	0.0842&	0.0388&	0.0377\\\hline
&$s_G$ averaged in $S'$ &0.911&	0.2632&	0.0274&	0.0344&	0.0041\\\hline
&$s_S$ averaged in $S'$ &1.158&	1.201&	1.072&	1.049&	0.992\\\hline\hline
T3&$s_G$ averaged in $G'$ &1.03&	1.057&	0.9868&	1.019&	1.009\\\hline
&$s_S$ averaged in $G'$ &0.869&	0.6141&	0.1164&	0.0085&	0.0067\\\hline
&$s_G$ averaged in $S'$ &1.001&	0.428&	0.0247&	0.0034&	0.0003\\\hline
&$s_S$ averaged in $S'$ &1.101&	1.177	&	0.947&	0.8359&	0.783\\\hline
\end{tabular}\end{center}
\end{table}
\begin{table}[htbp]
\caption{The number of falsified model points with respect to the integrated luminosity $L$.
Eq. (\ref{disc_cri}) is used for the falsification criteria.}
\label{falsified}
\begin{center}\begin{tabular}{|c|c|c|c|c|c|c|}
\hline
&$L$ & 1 fb$^{-1}$ & 3 fb$^{-1}$ & 10 fb$^{-1}$ & 30 fb$^{-1}$ & 100 fb$^{-1}$ \\\hline\hline
T1&$G'$ pts ($/1091$) falsified in $G$ & 0&	1&	2&	1&	4\\\hline
&$G'$ pts ($/1091$) falsified in $S$ &0&	323&	994&	1051&	1058\\\hline
&$S'$ pts ($/754$) falsified in $G$ &0&	126&	199&	176&	169\\\hline
&$S'$ pts ($/754$) falsified in $S$ &0&	0&	0&	5&	6\\\hline\hline
T2&$G'$ pts ($/1091$) falsified in $G$ & 0&	1&	2&	0&	0\\\hline
&$G'$ pts ($/1091$) falsified in $S$ &0&	394&	1049&	1089&	1089\\\hline
&$S'$ pts ($/754$) falsified in $G$ &10&	523&	753&	753&	751\\\hline
&$S'$ pts ($/754$) falsified in $S$ &0&	0&	0&	0&	0\\\hline\hline
T3&$G'$ pts ($/1091$) falsified in $G$ & 0&	0&	1&	0&	0\\\hline
&$G'$ pts ($/1091$) falsified in $S$ &0&	136&	921&	1091&	1091\\\hline
&$S'$ pts ($/754$) falsified in $G$ &0&	336&	752&	753&	754\\\hline
&$S'$ pts ($/754$) falsified in $S$ &0&	0&	0&	1&	1\\\hline
\end{tabular}\end{center}
\end{table}
First, we see that $s_M$ calculated in the corresponding model itself is consistent with unity.
There seems a tendency of slightly lower values in the high luminosity especially for the mSUGRA case in T3. This may be due to varying
errors of the significances by the statistical uncertainty of the Monte Carlo simulation. We checked this tendency
vanishes if we exclude this uncertainty in the error of the significances. (See Table \ref{woMCerr}.)
In contrast, $s_M$ calculated in the wrong model decreases with higher integrated luminosity,
indicating that more model points are discriminated from the other model.

This fact is more clear in Table \ref{falsified}, where
the number of falsified model points using the criterion in Eq.~(\ref{disc_cri}) is presented.
We see a large portion of the model points are falsified from the other model at higher luminosity while
most of them remain consistent within the home model.
Let us see each tests in more detail. In T1, many mSUGRA model points remain consistent
with being a mGMSB model even at 100 ${\rm fb}^{-1}$. This is because without using the tau-jet mode,
the slepton-NLSP GMSB model with nearly degenerated stau and the lightest neutralino may not be discriminated
from mSUGRA models as described in the previous section. These mGMSB model points conversely remain unfalsified as a mSUGRA model.
These models may be discriminated by using all five modes as in T3. The result shows all mGMSB model points are falsified from
being in the mSUGRA model set at 30 ${\rm fb}^{-1}$.
The T2 using combined significances also shows a good separation of models. In particular, it shows a better falsifiability
in lower luminosity. This is because T2 uses only three variables for the $\chi^2$ test and has an advantage in statistics.
In Table \ref{woMCerr}, we list the same result but without the Monte Carlo statistical error in the significances for T3.
\begin{table}[htbp]
\caption{Same as Tables \ref{sGsS} and \ref{falsified} for T3,
without the Monte Carlo statistical error included in the significances.}
\label{woMCerr}
\begin{center}\begin{tabular}{|c|c|c|c|c|c|c|}
\hline
&$L$ & 1 fb$^{-1}$ & 3 fb$^{-1}$ & 10 fb$^{-1}$ & 30 fb$^{-1}$ & 100 fb$^{-1}$ \\\hline\hline
T3&$s_G$ averaged in $G'$ &1.031&	1.053&	0.974&	1.009&	1.012\\\hline
&$s_S$ averaged in $G'$ &0.856&	0.631&	0.093&	0.007&	0\\\hline
&$s_G$ averaged in $S'$ &0.996&	0.394&	0.014&	0&	0\\\hline
&$s_S$ averaged in $S'$ &1.153&	1.277&	1.06&	1.041&	0.997\\\hline\hline
&$G'$ pts ($/1091$) falsified in $G$ & 0&	0&	1&	0&	0\\\hline
&$G'$ pts ($/1091$) falsified in $S$ &0	&137	&959&	1091&1091\\\hline
&$S'$ pts ($/754$) falsified in $G$ &0	&360&752	&754&	754\\\hline
&$S'$ pts ($/754$) falsified in $S$ &0&	0&	0&	0&	0\\\hline
\end{tabular}\end{center}
\end{table}
An important conclusion of this result is that even with the inclusion of the SM background and the effect of the detector efficiencies,
the model discrimination among the GMSB and SUGRA models using the method gives a very good result at the 30 ${\rm fb}^{-1}$, i.e.
at the time of the 5$\sigma$ (or more) discovery.

In the above result, the model points are generated on a flat grid in the parameter space and the tested models
$G'$ and $S'$ are subset of $G$ and $S$, respectively. There is a worry that the number of mis-falsified models
is underestimated and there might exist small characteristic regions dropped out from the grid.
To check this problem, we generate a new set of test model points randomly in the parameter space.
The used regions in the parameter space is $-100\leq\Lambda_s\leq400$ (TeV), $100\leq\Lambda_g\leq400$ (TeV) and
$5\leq{\rm tan}\,\beta\leq50$ for the mGMSB model and $0\leq m_0\leq1200$ (GeV), $100\leq m_{1/2}\leq600$ (GeV),
$5\leq{\rm tan}\,\beta\leq50$ and $-1\leq A_0/m_0\leq1$ for the mSUGRA model.
After requiring the experimental constraints and the discovery condition ($5\sigma$ at 30 ${\rm fb}^{-1}$), 
we generated 885 model points (denoted by $G''$)
for the mGMSB model and 1131 model points (denoted by $S''$) for the mSUGRA model.
We present the result of the discrimination test T3 in Table \ref{random_disc}.
\begin{table}[htbp]
\caption{The number of falsified model points by the significance test T3 for 
randomly chosen model points in $G''$ and $S''$.}
\label{random_disc}
\begin{center}\begin{tabular}{|c|c|c|c|c|c|c|}
\hline
&$L$ & 1 fb$^{-1}$ & 3 fb$^{-1}$ & 10 fb$^{-1}$ & 30 fb$^{-1}$ & 100 fb$^{-1}$ \\\hline
T3&$G''$ pts ($/885$) falsified in $G$ & 0&	3&	4&	4&	5\\\hline
&$G''$ pts ($/885$) falsified in $S$ &0	&133	&809&	885&	885\\\hline
&$S''$ pts ($/1131$) falsified in $G$ &0	&507&1129	&1131&	1131\\\hline
&$S''$ pts ($/1131$) falsified in $S$ &0&	0&	0&	0&	0\\\hline
\end{tabular}\end{center}
\end{table}
The result is essentially same as the previous result for $G'$ and $S'$ and showing the underestimation is very small.

\section{Discussion and conclusion}\label{conclusion}
In this paper, we discussed a strategy of model discrimination between the mSUGRA and the mGMSB models at a very early stage of the LHC 
by comparing significances in multiple search modes.
We defined the difference of the LHC signatures among model points in a quantitative form and
show that the discrimination is possible with a relatively low integrated luminosity.
Our estimation of the significances is conservative and the discrimination may be better in the actual experiments.
This model discrimination is very important for both model-building and cosmology, since
they predict completely different cosmological history and the dark matter.
We used a relatively small number of modes in this paper.
Although such a coarse graining of LHC signals gives only coarse information,
it can discriminate the models qualitatively 
and  guides to research at a much higher integrated luminosity.
Though in this paper, we focused on only discrimination between  the mSUGRA and mGMSB models,
our method can be applied for more generic cases.
For example, we can also discuss the discrimination within the mSUGRA models.
The so-called co-annihlation and focus regions, which are favored from the viewpoint of dark matter,
can be separated by using similar method. 
This method makes it possible to compare the LHC signals of generic BSM models qualitatively.

\subsection*{Acknowledgement}
We would like to thank K. Hamaguchi and T. T. Yanagida for useful discussions.
The work of E.N. is supported in part by JSPS Research Fellowships for Young Scientists.

\appendix
\section{Supplemental data of example model points}
In this appendix, we collect supplemental data of our example model points in Tab. \ref{ex_pts}
for convenience.
In Tab. \ref{tab_mass} and \ref{tab_mass_2}, we list up the mass spectrum of the superparticles of the example model points
together with their dominant decay channels.

\begin{table}[htbp]
  \centering
  \caption{Superparticles' masses (in unit of GeV) and their dominant decay channels for the example model points
PT 1 to 6 (mGMSB model) in Tab. \ref{ex_pts}. From top to bottom, the particles' label represent gluino,
down-type squarks, stop, sbottom, selectron, stau, charginos and neutralinos. The gravitino is denoted by $\tilde{G}$.}
  \label{tab_mass}
  \begin{tabular}{|c||r|l||r|l||r|l|} 
    \hline \hline
& PT 1 & Decay mode & PT 2 & Decay mode & PT 3 & Decay mode \\ \hline
$\tilde{g}$ & 1530.7 & $\tilde{\chi}_1^\pm q\bar{q}',\tilde{\chi}_2^0q\bar{q},\tilde{\chi}_1^0q\bar{q}$ & 1530.1 & $\tilde{\chi}_1^\pm q\bar{q}',\tilde{\chi}_2^0q\bar{q},\tilde{\chi}_1^0q\bar{q}$ & 1319.4 & $\tilde{q}q$ \\
$\tilde{d}_{L}$ & 2546.9 & $\tilde{g}d,\tilde{\chi}_1^-u$ & 2546.3 & $\tilde{g}d,\tilde{\chi}_1^-u$ & 1306.4 & $\tilde{\chi}_1^-u,\tilde{\chi}_2^-u$ \\
$\tilde{d}_{R}$ & 2414.7 & $\tilde{g}d$ & 2414.0 & $\tilde{g}d$ & 1252.4 & $\tilde{\chi}_1^0d$ \\
$\tilde{t}_{2}$ & 2472.2 & $\tilde{g}t,\tilde{\chi}_1^+b$ & 2441.1 & $\tilde{g}t,\tilde{\chi}_1^+b$ & 1277.2 & $\tilde{\chi}_4^0t,\tilde{\chi}_3^0t,\tilde{\chi}_2^+b$ \\
$\tilde{t}_{1}$ & 2208.4 & $\tilde{g}t,\tilde{\chi}_2^+b$ & 2210.9 & $\tilde{g}t,\tilde{\chi}_2^+b$ & 1147.8 & $\tilde{\chi}_1^+b,\tilde{\chi}_3^0t,\tilde{\chi}_2^+b$ \\
$\tilde{b}_{2}$ & 2454.8 & $\tilde{g}b,\tilde{\chi}_2^-t$ & 2428.0 & $\tilde{g}b,\tilde{\chi}_2^-t$ & 1261.0 & $\tilde{\chi}_2^-t,\tilde{\chi}_1^-t$ \\
$\tilde{b}_{1}$ & 2409.2 & $\tilde{g}b$ & 2344.6 & $\tilde{g}b$ & 1247.0 & $\tilde{\chi}_1^-t,\tilde{\chi}_2^-t$ \\
$\tilde{e}_{L}$ & 851.1 & $\tilde{\chi}_1^-\nu,\tilde{\chi}_2^0e,\tilde{\chi}_1^0e$ & 850.7 & $\tilde{\chi}_1^-\nu,\tilde{\chi}_2^0e,\tilde{\chi}_1^0e$ & 394.4 & $\tilde{\chi}_1^0e$ \\
$\tilde{e}_{R}$ & 415.7 & $\tilde{\chi}_1^0e$ & 416.9 & $\tilde{\chi}_1^0e$ & 188.7 & $\tilde{G}e$ \\
$\tilde{\tau}_{2}$ & 848.2 & $\tilde{\chi}_1^-\nu,\tilde{\chi}_2^0\tau,\tilde{\chi}_1^0\tau$ & 841.8 & $\tilde{\chi}_1^-\nu,\tilde{\chi}_2^0\tau,\tilde{\chi}_1^0\tau$ & 392.1 & $\tilde{\chi}_1^0\tau$ \\
$\tilde{\tau}_{1}$ & 422.0 & $\tilde{\chi}_1^0\tau$ & 367.4 & $\tilde{\chi}_1^0\tau$ & 191.3 & $\tilde{G}\tau$ \\
$\tilde{\chi}^+_{2}$ & 877.7 & $\tilde{\chi}_1^+h/Z,\tilde{\chi}_2^0W^+$ & 860.8 & $\tilde{\chi}_1^+h/Z,\tilde{\chi}_2^0W^+$ & 537.7 & $\tilde{\ell}\nu,\tilde{\nu}\ell$ \\
$\tilde{\chi}^+_{1}$ & 529.8 & $\tilde{\chi}_1^0W^+$ & 531.3 & $\tilde{\tau}\nu,\tilde{\chi}_1^0W^+$ & 423.8 & $\tilde{\chi}_1^0W^+,\tilde{\nu}\ell$ \\
$\tilde{\chi}^0_{4}$ & 877.3 & $\tilde{\chi}_1^+W^-,\tilde{\chi}_2^0h$ & 860.2 & $\tilde{\chi}_1^+W^-,\tilde{\chi}_2^0h$ & 539.8 & $\tilde{\ell}\ell,\tilde{\nu}\nu$ \\
$\tilde{\chi}^0_{3}$ & 867.6 & $\tilde{\chi}_1^+W^-,\tilde{\chi}_2^0Z$ & 852.0 & $\tilde{\chi}_1^+W^-,\tilde{\chi}_2^0Z$ & 483.0 & $\tilde{\chi}_1^0Z,\tilde{\ell}\ell$ \\
$\tilde{\chi}^0_{2}$ & 529.5 & $\tilde{\chi}_1^0h$ & 530.8 & $\tilde{\tau}\tau,\tilde{\chi}_1^0h$ & 422.2 & $\tilde{\chi}_1^0h,\tilde{\ell}\ell$ \\
$\tilde{\chi}^0_{1}$ & 273.4 & $\tilde{G}\gamma/Z$ & 273.8 & $\tilde{G}\gamma/Z$ & 240.6 & $\tilde{\ell}_R\ell,\tilde{\tau}_1\tau$ \\
    \hline \hline
& PT 4 & Decay mode & PT 5 & Decay mode & PT 6 & Decay mode \\ \hline
$\tilde{g}$ & 941.5 & $\tilde{\chi}_2^0q\bar{q},\tilde{\chi}_1^0q\bar{q},\tilde{\chi}_1^\pm q\bar{q}'$ & 1770.1 & $\tilde{q}q$ & 1595.6 & $\tilde{q}q$ \\
$\tilde{d}_{L}$ & 1192.3 & $\tilde{g}d,\tilde{\chi}_1^-u$ & 1253.0 & $\tilde{\chi}_2^-u,\tilde{\chi}_4^0d$ & 1262.1 & $\tilde{\chi}_2^-u,\tilde{\chi}_4^0d$ \\
$\tilde{d}_{R}$ & 1137.5 & $\tilde{g}d$ & 1218.1 & $\tilde{\chi}_1^0d$ & 1220.2 & $\tilde{\chi}_1^0d$ \\
$\tilde{t}_{2}$ & 1146.5 & $\tilde{\chi}_1^+b,\tilde{\chi}_4^0t,\tilde{\chi}_3^0t$ & 1242.6 & $\tilde{\chi}_3^0t,\tilde{\chi}_2^0t,\tilde{\chi}_2^+b$ & 1230.4 & $\tilde{\chi}_2^+b,\tilde{\chi}_3^0t,\tilde{\chi}_2^0t$ \\
$\tilde{t}_{1}$ & 1042.6 & $\tilde{\chi}_2^+b,\tilde{\chi}_3^0t$ & 1120.1 & $\tilde{\chi}_1^+b,\tilde{\chi}_3^0t,\tilde{\chi}_2^0t$ & 1122.6 & $\tilde{\chi}_1^+b,\tilde{\chi}_3^0t,\tilde{\chi}_2^0t$ \\
$\tilde{b}_{2}$ & 1139.4 & $\tilde{\chi}_2^-t,\tilde{g}b$ & 1220.5 & $\tilde{\chi}_1^-t,\tilde{\chi}_2^-t$ & 1218.6 & $\tilde{\chi}_1^-t,\tilde{\chi}_2^-t$ \\
$\tilde{b}_{1}$ & 1093.5 & $\tilde{g}b,\tilde{\chi}_1^-t$ & 1209.5 & $\tilde{\chi}_1^-t,\tilde{\chi}_2^-t$ & 1178.0 & $\tilde{\chi}_1^-t,\tilde{\chi}_2^0b$ \\
$\tilde{e}_{L}$ & 378.7 & $\tilde{\chi}_1^-\nu,\tilde{\chi}_1^0e,\tilde{\chi}_2^0e$ & 327.3 & $\tilde{\tau}_1\tau e$ & 350.1 & $\tilde{\chi}_1^0e$ \\
$\tilde{e}_{R}$ & 184.9 & $\tilde{\chi}_1^0e$ & 146.7 & $\tilde{G}e$ & 163.0 & $\tilde{\tau}\tau e$ \\
$\tilde{\tau}_{2}$ & 383.5 & $\tilde{\tau}_1Z/h,\tilde{\chi}_1^0\tau$ & 322.3 & $\tilde{\tau}_1Z/h$ & 354.9 & $\tilde{\tau}_1Z/h$ \\
$\tilde{\tau}_{1}$ & 137.5 & $\tilde{G}\tau$ & 149.5 & $\tilde{G}\tau$ & 114.1 & $\tilde{G}\tau$ \\
$\tilde{\chi}^+_{2}$ & 473.0 & $\tilde{\chi}_2^0W^+,\tilde{\chi}_1^+Z$ & 688.2 & $\tilde{\ell}\nu,\tilde{\nu}\ell$ & 616.8 & $\tilde{\ell}\nu,\tilde{\nu}\ell$ \\
$\tilde{\chi}^+_{1}$ & 304.8 & $\tilde{\tau}\nu$ & 445.2 & $\tilde{\nu}\ell,\tilde{\chi}_1^0W^+$ & 441.5 & $\tilde{\tau}_1\nu,\tilde{\nu}\tau$ \\
$\tilde{\chi}^0_{4}$ & 472.9 & $\tilde{\chi}_1^+W^-,\tilde{\chi}_2^0h$ & 695.3 & $\tilde{\nu}\nu,\tilde{\ell}_L\ell$ & 623.1 & $\tilde{\ell}\ell,\tilde{\nu}\nu$ \\
$\tilde{\chi}^0_{3}$ & 453.7 & $\tilde{\chi}_1^+W^-,\tilde{\tau}_1\tau,\tilde{\chi}_2^0Z$ & 454.0 & $\tilde{\chi}_1^0Z,\tilde{\tau}_1\tau$ & 454.6 & $\tilde{\tau}_1\tau$ \\
$\tilde{\chi}^0_{2}$ & 304.1 & $\tilde{\tau}_1\tau$ & 447.3 & $\tilde{\tau}_1\tau,\tilde{\ell}_R\ell$ & 439.4 & $\tilde{\tau}_1\tau$ \\
$\tilde{\chi}^0_{1}$ & 160.9 & $\tilde{\tau}_1\tau$ & 342.8 & $\tilde{\tau}_1\tau,\tilde{\ell}_R\ell$ & 306.3 & $\tilde{\tau}_1\tau,\tilde{\ell}_R\ell$ \\
    \hline \hline
  \end{tabular}
\end{table}
\begin{table}[htbp]
  \centering
  \caption{Same as Tab. \ref{tab_mass}, but for example model points
PT 7 to 10 (mSUGRA model) in Tab. \ref{ex_pts}.
Since the list of decay modes is very similar for PT 9 and PT 10, their lists are gathered.}
  \label{tab_mass_2}
  \begin{tabular}{|c||r|l||r|l||r|r|l|} 
    \hline \hline
& PT 7 & Decay mode & PT 8 & Decay mode & PT 9 & PT 10 & Decay mode \\ \hline
$\tilde{g}$ & 1157.2 & $\tilde{q}q$ & 1156.7 & $\tilde{q}q$ & 689.7 & 689.1 & $\tilde{\chi}_1^\pm q\bar{q}',\tilde{\chi}_2^0q\bar{q},\tilde{\chi}_1^0q\bar{q}$ \\
$\tilde{d}_{L}$ & 1086.1 & $\tilde{\chi}_1^-u,\tilde{\chi}_2^0d$ & 1086.8 & $\tilde{\chi}_1^-u,\tilde{\chi}_2^0d$ & 1223.3 & 1224.2 & $\tilde{g}d,\tilde{\chi}_1^-u$ \\
$\tilde{d}_{R}$ & 1043.4 & $\tilde{\chi}_1^0d$ & 1044.0 & $\tilde{\chi}_1^0d$ & 1217.3 & 1218.0 & $\tilde{g}d$ \\
$\tilde{t}_{2}$ & 1035.7 & $\tilde{\chi}_1^+b,\tilde{\chi}_4^0t$ & 1008.4 & $\tilde{\chi}_1^+b,\tilde{\chi}_4^0t$ & 1031.5 & 961.8 & $\tilde{g}t,\tilde{\chi}_1^+b$ \\
$\tilde{t}_{1}$ & 813.5 & $\tilde{\chi}_1^+b,\tilde{\chi}_1^0t,\tilde{\chi}_2^+b$ & 816.1 & $\tilde{\chi}_1^+b,\tilde{\chi}_1^0t,\tilde{\chi}_2^+b$ & 755.4 & 761.9 & $\tilde{\chi}_2^+b,\tilde{\chi}_3^0t$ \\
$\tilde{b}_{2}$ & 1038.5 & $\tilde{\chi}_2^-t,\tilde{\chi}_1^0b$ & 1000.3 & $\tilde{\chi}_2^-t$ & 1204.5 & 1076.4 & $\tilde{g}b$ \\
$\tilde{b}_{1}$ & 994.7 & $\tilde{\chi}_1^-t,\tilde{\chi}_2^-t,\tilde{\chi}_2^0b$ & 940.0 & $\tilde{\chi}_1^-t,\tilde{\chi}_2^-t,\tilde{\chi}_2^0b$ & 1016.5 & 941.1 & $\tilde{g}b,\tilde{\chi}_2^-t$ \\
$\tilde{e}_{L}$ & 427.9 & $\tilde{\chi}_1^0e,\tilde{\chi}_1^-\nu$ & 427.9 & $\tilde{\chi}_1^0e,\tilde{\chi}_1^-\nu$ & 1111.7 & 1112.0 & $\tilde{\chi}_1^-\nu,\tilde{\chi}_2^0e$ \\
$\tilde{e}_{R}$ & 321.2 & $\tilde{\chi}_1^0e$ & 321.6 & $\tilde{\chi}_1^0e$ & 1103.7 & 1103.7 & $\tilde{\chi}_1^0e$ \\
$\tilde{\tau}_{2}$ & 426.9 & $\tilde{\chi}_1^0\tau,\tilde{\chi}_1^-\nu$ & 423.9 & $\tilde{\tau}_1Z/h,\tilde{\chi}_1^0\tau$ & 1107.5 & 1027.9 & $\tilde{\chi}_1^-\nu,\tilde{\chi}_2^0\tau$ \\
$\tilde{\tau}_{1}$ & 316.3 & $\tilde{\chi}_1^0\tau$ & 213.6 & $\tilde{\chi}_1^0\tau$ & 1092.4 & 920.4 & $\tilde{\chi}_1^0\tau$ \\
$\tilde{\chi}^+_{2}$ & 650.5 & $\tilde{\chi}_2^0W^+,\tilde{\chi}_1^+Z/h$ & 637.0 & $\tilde{\chi}_2^0W^+,\tilde{\chi}_1^+Z/h$ & 438.0 & 404.0 & $\tilde{\chi}_2^0W^+,\tilde{\chi}_1^+Z/h$ \\
$\tilde{\chi}^+_{1}$ & 388.4 & $\tilde{\chi}_1^0W^+,\tilde{\tau}\nu$ & 390.6 & $\tilde{\tau}_1\nu$ & 200.4 & 201.1 & $\tilde{\chi}_1^0W^+$ \\
$\tilde{\chi}^0_{4}$ & 651.0 & $\tilde{\chi}_1^+W^-,\tilde{\chi}_2^0h$ & 636.8 & $\tilde{\chi}_1^+W^-,\tilde{\chi}_2^0h$ & 438.1 & 403.4 & $\tilde{\chi}_1^+W^-,\tilde{\chi}_2^0h$ \\
$\tilde{\chi}^0_{3}$ & 636.4 & $\tilde{\chi}_1^+W^-,\tilde{\chi}_2^0Z$ & 624.2 & $\tilde{\chi}_1^+W^-,\tilde{\chi}_2^0Z$ & 422.3 & 388.4 & $\tilde{\chi}_1^+W^-,\tilde{\chi}_2^0Z$ \\
$\tilde{\chi}^0_{2}$ & 387.9 & $\tilde{\chi}_1^0h,\tilde{\tau}_1\tau$ & 390.0 & $\tilde{\tau}_1\tau$ & 199.8 & 200.4 & $\tilde{\chi}_1^0Z$ \\
$\tilde{\chi}^0_{1}$ & 205.7 & stable & 206.3 & stable & 105.8 & 106.4 & stable \\
    \hline \hline
  \end{tabular}
\end{table}

In Tab. \ref{NsNb}, we list the number of signal events for the example model
points and the number of the background events for the search modes in
Tab. \ref{cuts}. For the four jets with no lepton mode, the showed results are 
event numbers under the hardest cut in Tab. \ref{cuts}, which optimizes the significance
for most of the model points.

\begin{table}[htbp]
\caption{The number of signal events of the example model points for the five modes in Tab. \ref{cuts}
together with our estimation of the SM background. The event numbers are normalized for an integrated
luminosity of 30 ${\rm fb}^{-1}$ at the 7 TeV run.}
\label{NsNb}
\begin{center}\begin{tabular}{|c|c|c|c|c|c|}
\hline
 & 0$\ell+2$ jets  & 0$\ell+4$ jets & 2$\gamma$ & $2\tau$ & SS $2\ell$\\ \hline\hline
PT 1 & 0.3 & 0.5 &45.3 & 0 & 0\\ \hline
PT 2 & 0 & 0.3 &44.9 & 0.17 & 0\\ \hline
PT 3 & 9.1 & 6.0 &0 & 6.04 & 63.5\\ \hline
PT 4 & 67.4 & 85.3 &0 & 65.1 & 10.1\\ \hline
PT 5 & 0 & 10.7 &0 & 10.7 & 57.8\\ \hline
PT 6 & 13.5 & 9.0 &0 & 13.5 & 57.5\\ \hline
PT 7 & 310.4 & 79.0 &1.1 & 1.14 & 10.3\\ \hline
PT 8 & 333.8 & 55.8 &0 & 2.37 & 3.55\\ \hline
PT 9 & 71.1 & 94.0 &0 & 2.30 & 34.4\\ \hline
PT 10 & 51.84 & 78.9 &0 & 9.01 & 37.2\\ \hline\hline
SM BKG & 195.1 & 44.9 &7.68 & 14.8 & 20.6\\ \hline
\end{tabular}\end{center}
\end{table}

\end{document}